\documentclass[sn-mathphys-num]{sn-jnl}

\usepackage{graphicx}%
\usepackage{multirow}%
\usepackage{amsmath,amssymb,amsfonts}%
\usepackage{amsthm}%
\usepackage{mathrsfs}%
\usepackage[title]{appendix}%
\usepackage{xcolor}%
\usepackage{textcomp}%
\usepackage{manyfoot}%
\usepackage{booktabs}%
\usepackage{algorithm}%
\usepackage{algorithmicx}%
\usepackage{algpseudocode}%
\usepackage{listings}%
\usepackage{xspace}
\usepackage[inline]{enumitem}
\usepackage{colortbl}
\usepackage{tabularx}
\usepackage{forest}
\usepackage{tikz}
\usetikzlibrary{arrows.meta,arrows}

\newcommand{\name}{Perimeterless\xspace}

\newcommand{\app}{Airline Booking\xspace}

\newcommand{\users}{application developers\xspace}
\newcommand{\etal}{\textit{et al.}\xspace}

\begin{document}

\title[Article Title]{Observability and Incident Response in Managed Serverless Environments Using Ontology-Based Log Monitoring}


\author[1]{\fnm{Lavi} \sur{Ben-Shimol}}
\author[1]{\fnm{Edita} \sur{Grolman}}
\author[1]{\fnm{Aviad} \sur{Elyashar}}
\author[1]{\fnm{Inbar} \sur{Maimon}}
\author[1]{\fnm{Dudu} \sur{Mimran}}
\author[1]{\fnm{Oleg} \sur{Brodt}}
\author[2]{\fnm{Martin} \sur{Strassmann}}
\author[2]{\fnm{Heiko} \sur{Lehmann}}
\author[1]{\fnm{Yuval} \sur{Elovici}}
\author*[1]{\fnm{Asaf} \sur{Shabtai}}\email{shabtaia@bgu.ac.il}

\affil*[1]{\orgdiv{Department of Software and Information Systems Engineering}, \orgname{Ben-Gurion University of the Negev}, \country{Israel}}

\affil[2]{\orgname{Deutsche Telekom AG}}


\abstract{In a fully managed serverless environment, the cloud service provider is responsible for securing the cloud infrastructure, thereby reducing the operational and maintenance efforts of application developers.
However, this environment limits the use of existing cybersecurity frameworks and tools, which reduces observability and situational awareness capabilities (e.g., risk assessment, incident response).
In addition, existing security frameworks for serverless applications do not generalize well to all application architectures and usually require adaptation, specialized expertise, etc. for use in fully managed serverless environments.
In this paper, we introduce a three-layer security scheme for applications deployed in fully managed serverless environments.
The first two layers involve a unique ontology based solely on serverless logs which is used to transform them into a unified application activity knowledge graph.
In the third layer, we address the need for observability and situational awareness capabilities by implementing two situational awareness tools that utilizes the graph-based representation: 1) An incident response dashboard that leverages the ontology to visualize and examine application activity logs in the context of cybersecurity alerts.
Our user study showed that the dashboard enabled participants to respond more accurately and quickly to new security alerts than the baseline tool. 
2) A criticality of asset (CoA) risk assessment framework that enables efficient expert-based prioritization in cybersecurity contexts.}




\maketitle

\section{Introduction}\label{sec:intro}
In recent years, serverless computing in cloud environments has seen rapid growth. 
This model offers significant advantages, such as reduced operational effort, efficient scaling, and cost-effective pay-as-you-go models for application developers~\cite{marketvalue2021,parast2022cloud}.
Serverless architecture decomposes applications into atomic, stateless function units, while cloud service providers (CSPs) provide fully managed resources like storage and queue management~\cite{leitner2019mixed}.
These features are particularly beneficial for streaming, synchronization, batch tasks, and API applications~\cite{redhat_what_is,eismann2021state,castro2019rise}.
However, fully managed serverless environments require that organizations and \users use a shared security model, which presents new cybersecurity challenges~\cite{shafiei2022serverless}.
In this model, the CSP is responsible for securing the cloud infrastructure, platforms provided, network traffic, etc., while \users are responsible for securing the application's data and functionalities.

\begin{figure*}[h]
    \centering
        \includegraphics[trim=0cm 0cm 0cm 0cm, clip,width=0.99\textwidth]
        {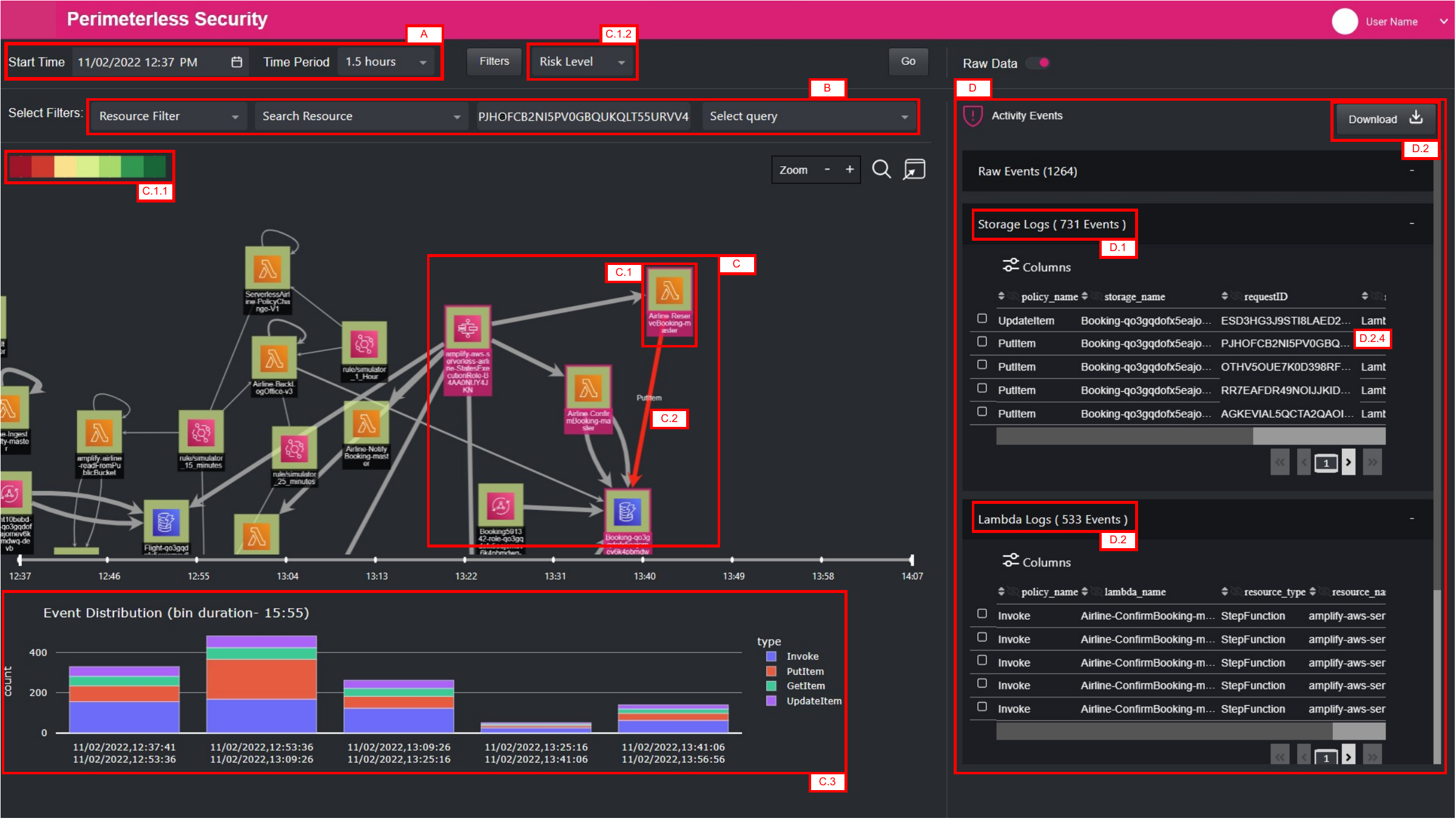}
    \caption{\name incident response dashboard.}
    \label{fig:dashboard_image}
\end{figure*}

According to the Cloud Security Alliance\footnote{\href{https://cloudsecurityalliance.org/artifacts/the-12-most-critical-risks-for-serverless-applications}{https://cloudsecurityalliance.org/artifacts/\\the-12-most-critical-risks-for-serverless-applications}}
and Netwrix cloud security survey,\footnote{\url{https://www.netwrix.com/2021_cloud_data_security_report.html}} 
serverless application monitoring challenges include the following:
1) While beneficial in certain scenarios, the high granularity of the CSP logging capabilities can make identifying and mitigating application-layer attacks, e.g., SQL injection, and brute-force authentication attempts more challenging.
2) Due to the multiple triggering options (e.g., URL requests, APIs, emails, rules), serverless apps provide obscure context and limited root cause analysis capabilities.
3) Each serverless service provided by a CSP is fully managed; therefore \users cannot implement and use many of their existing cybersecurity frameworks and tools, making it difficult to evaluate the application's attack surface.
4) Cybersecurity personnel often lack the necessary knowledge and expertise in managed serverless environments, limiting their ability to detect, prioritize, and investigate incidents~\cite{marin2022serverless}.

While cybersecurity solutions are often tailored for traditional, unmanaged environments, they are generally inadequate in fully managed serverless environments~\cite {li2022securing}. 
Existing security frameworks for serverless applications~\cite{datta2020valve,alpernas2018secure,sankaran2020workflow,jegan2020guarding,polinsky2021sciffs,obetz2019static,mendsaikhan2020quantifying,banse2021cloud} do not generalize well to all application architectures and usually require adaptation, expertise, etc.
Specifically, these frameworks often require specialized knowledge and access to the application code, and may introduce performance overhead.

In light of these complexities and challenges, robust cybersecurity situational awareness (CSA) capabilities are critical for enhancing observability and minimizing the mean time to detect (MTTD) in cybersecurity incidents~\cite{marin2022serverless,zimmerman2014ten,brownlee1998expectations}.
To gain effective CSA, security teams must perform tasks such as real-time monitoring, incident response, and risk assessment~\cite{brownlee1998expectations,zimmerman2014ten}.
These tasks, which reflect CSA capabilities, often rely on audit log analysis~\cite{cir} and assisting in attack detection and post-incident actions~\cite{kim2022vulnerabilities,OWASP,cyberwire,darkreading}. 
To improve CSA capabilities, in this paper, we propose the \emph{\name} stack, a novel approach that relies on a graph-based methodology and a unique log-based ontology and enables audit logs to be transformed into contextualized graphs.

The \name stack introduces a novel three-layer security scheme designed for fully managed serverless environments, founded on a zero trust architecture principle that advocates for no inherent trust within or outside organizational perimeters~\cite{rose2020zero}. 
At the core of this scheme, we have developed an innovative ontology based on serverless application logs. This ontology facilitates the transformation of these logs into a cohesive application activity knowledge graph, which forms the foundation of the first two layers of our security scheme.
Building on this foundation, the \name stack's third layer innovatively addresses the challenges of observability and situational awareness through the introduction of two key situational awareness tools.
The first tool is the incident response (IR) dashboard (Figure~\ref{fig:dashboard_image}), relies on the ontology to provide graph-based visualization and analysis of application activity logs in the context of cybersecurity alerts, significantly enhancing the speed and accuracy of incident response.
The second tool, a criticality of asset (CoA) risk assessment framework, leverages the same ontology to offer expert-based prioritization of assets, thus improving resource management and application observability within cybersecurity contexts.

Our proposed \name stack has several key benefits:
\begin{enumerate*}
    \item It does not require knowledge regarding the application's resources or code. 
    \item It is based on standard application API logs and thus does not require access to the application's data itself.
    \item It is designed as a standalone solution that does not require integration in the application; therefore, it does not add any additional overhead to the application's operations.
\end{enumerate*}

We conducted a comprehensive evaluation of the proposed \name stack and CSA tools. 
This included the implementation of a proof of concept (PoC) using real-world log data from a serverless application deployed on Amazon Web Services (AWS).
To collect serverless application activity logs, we created a testbed to simulate benign and malicious activities.
To evaluate the \name IR dashboard tool, we performed a user study with 39 participants, in which we assessed the dashboard's effectiveness in terms of response time.
The results show a significant reduction in the MTTD for the study's participants whose ability to accurately investigate security improved over the baseline (with a maximum improvement of 18\% seen).
The decision-making approach of Saaty et al. served as a foundation~\cite{ahp_paper} for the development of our novel CoA ranking framework for resource prioritization; in our evaluation of the proposed ranking framework a strong Kendall-W agreement score of 0.71 was obtained~\cite{kendall1948rank,kokol2023exact,bukovvsek2024exact}.

This paper makes the following contributions:
\begin{enumerate}[topsep=0pt,noitemsep,leftmargin=*]
    \item We introduce a graph-based IR dashboard that leverages our proposed \name stack to visualize serverless applications' activity. 
    This enhanced visualization enables CSA tasks (e.g., incident response) to be performed more quickly and efficiently.
    \item The proposed CoA risk assessment framework improves application observability and enables the prioritization of application resource alerts.
    \item The \name stack enables security teams to develop various situational awareness tools, such as the IR dashboard and the CoA risk assessment framework, as demonstrated in this paper.
    \item We created a novel dataset of AWS serverless application logs, which is publicly available on GitHub (this will be published upon the paper's publication).
\end{enumerate}

\section{\label{sec:background}Background}
\subsection{Serverless Environment}
Serverless computing encompasses services that enable developers to build applications without the need to manage the underlying infrastructure.
Key categories include:
\begin{itemize}
    \item \textbf{Serverless functions} - Managed compute executions (e.g., AWS Lambda) that scale automatically without server management by developers~\cite{castro2017serverless}.
    \item \textbf{Serverless storage resources} - Scalable storage options (e.g., Amazon S3, Azure Blob Storage) offering high availability without managing physical storage~\cite{naumenko2021analysis}.
    \item \textbf{Application integration services} - Components (e.g., AWS Step Functions, Azure Logic Apps) for orchestrating serverless workflows and managing application interactions~\cite{eismann2021state}.
    \item \textbf{Identity and Access Management (IAM)} - Frameworks controlling permissions and access within the serverless environment, ensuring secure resource utilization~\cite{mohammed2019cloud,batra2019deploying}.
\end{itemize}

\subsection{Log Auditing in Fully Managed Serverless Environments}\label{background_logging}
In a CSP serverless environment, automatic log auditing capabilities are provided by the CSP.
In our research, we focused on the AWS environment, as it is the leading CSP today~\cite{statista2023aws}.
We created a testbed application (see Section~\ref{sec:demoapp}) and employed AWS CloudTrail to monitor and log API activity, which enabled activity and operational auditing~\cite{AWScloudtrail}.
We focus on monitoring the following services, since they are considered core serverless application resources:\\
\begin{itemize}
    \item \textbf{Lambda} - The AWS serverless function service produces three types of logs: function, extension, and platform logs. 
    Function and extension logs, which produce code-generated logs and access to extension log mechanisms, must be set by the application developer.
    Platform logs provide the code runtime logs (running duration, billed time, memory use, data transferred, etc).
    \item \textbf{Storage} - The data access of all AWS users and resources can be recorded for the serverless storage services, simple storage service (S3), and DynamoDB (NoSQL database).
\end{itemize}
As part of the API calls auditing, the logging for the application integration components is also collected using CloudTrail, providing the trigger information in the API event record.
These components have a variety of triggering options.
For example, upon the completion of an application flow, such as finalizing an order, the system can trigger an email notification, enqueue a new message, or initiate other processes.
These serverless application integration services include:
\begin{enumerate*}
    \item \emph{EventBridge} - a serverless bus event that is used to trigger event-driven functions (e.g., trigger Lambda functions on timed events)~\cite{event_bridge};
    \item \emph{API Gateway} - a service that enables the maintenance of RESTful APIs~\cite{api_gateway}; and 
    \item \emph{Step Function} - a workflow service that is used to orchestrate microservices, build pipelines using AWS services, etc.~\cite{stepfunction}.
\end{enumerate*}

\begin{figure}[h]
    \centering
    \begin{lstlisting}[basicstyle=\ttfamily\small, frame=single]
      "eventSource": <service_name>,
      "eventName": <operation>,
      "date": <date>
      "awsRegion": <region>,
      "resources": [{"type": "<resource type>",
                    "ARN": "<resource>"}],
      "userIdentity": {
        "type": <IAM-type>,
        "principalId": <principal-id>,
        "arn": "<aws resource name>"
      }
      "requestParameters": { .. }
    \end{lstlisting}
    \caption{AWS API log entry.}
    \label{fig:aws_api_log}
\end{figure}

Figure~\ref{fig:aws_api_log} presents a log entry template for an API request in an AWS serverless environment.
In this log, \texttt{eventSource} indicates the service type receiving the action, \texttt{eventName} specifies the API request's name, \texttt{resources} identifies the AWS resource executing the request, and \texttt{userIdentity} represents the credentials used.

\subsection{Cybersecurity Attacks in Cloud Applications}
\textbf{Denial-of-Wallet Attack}\label{background_dow}
Similar to denial-of-service (DoS) attacks, denial-of-wallet (DoW) attacks disrupt application availability, but they target the serverless pay-as-you-go model by draining the user's budget through excessive API calls\cite{haworth2020denialofwallet, kelly2021denial, claburn2021denialofwallet}.
Recognized as a significant concern by both the Open Worldwide Application Security Project (OWASP) and the Cyber Security Alliance, this attack vector underscores the need for effective countermeasures~\cite{nsrav2022dos,csa2022toptthreat9,shen2022gringotts}.

\textbf{Data Leakage}\label{background_dl}
Data leakage refers to the unintended dissemination of sensitive information to unauthorized parties, such as intellectual property, financial records, and patient details~\cite{shabtai2012survey}. 
Thales' 2023 research on cloud security found that 39\% of businesses experienced a data breach in the last year\cite{Thales2023}.
Data leakage attacks represent a significant risk to organizations and can have immediate and long-term consequences.
In such attacks, the attacker obtains sensitive information from the cloud environment.
Data leakage attacks can be performed by discovering and exploiting insecure serverless storage or by employing more comprehensive attack strategies to exploit various environments.
Several incidents in recent years have highlighted the risks of data leakage due to misconfigured Amazon S3 buckets~\cite{PortSwigger2022,Paganini2022,Eddy2022}.

\section{Testbed Application}
\label{sec:demoapp}
In this section, we introduce the Airline Booking serverless application,\footnote{\url{https://github.com/aws-samples/aws-serverless-airline-booking}} deployed on AWS, which serves as a testbed for both explaining and demonstrating our proposed \name stack, and for evaluating it.
This application is well-regarded in the literature and frequently utilized as a benchmark for research~\cite{EISMANN2022111294}, making it an ideal candidate for our purposes. 
We also describe our simulator that can generate activity logs for demonstrating how the \name stack functions within a real-world serverless application scenario.

\begin{figure}[h]
    \centering
\includegraphics[width=\columnwidth]{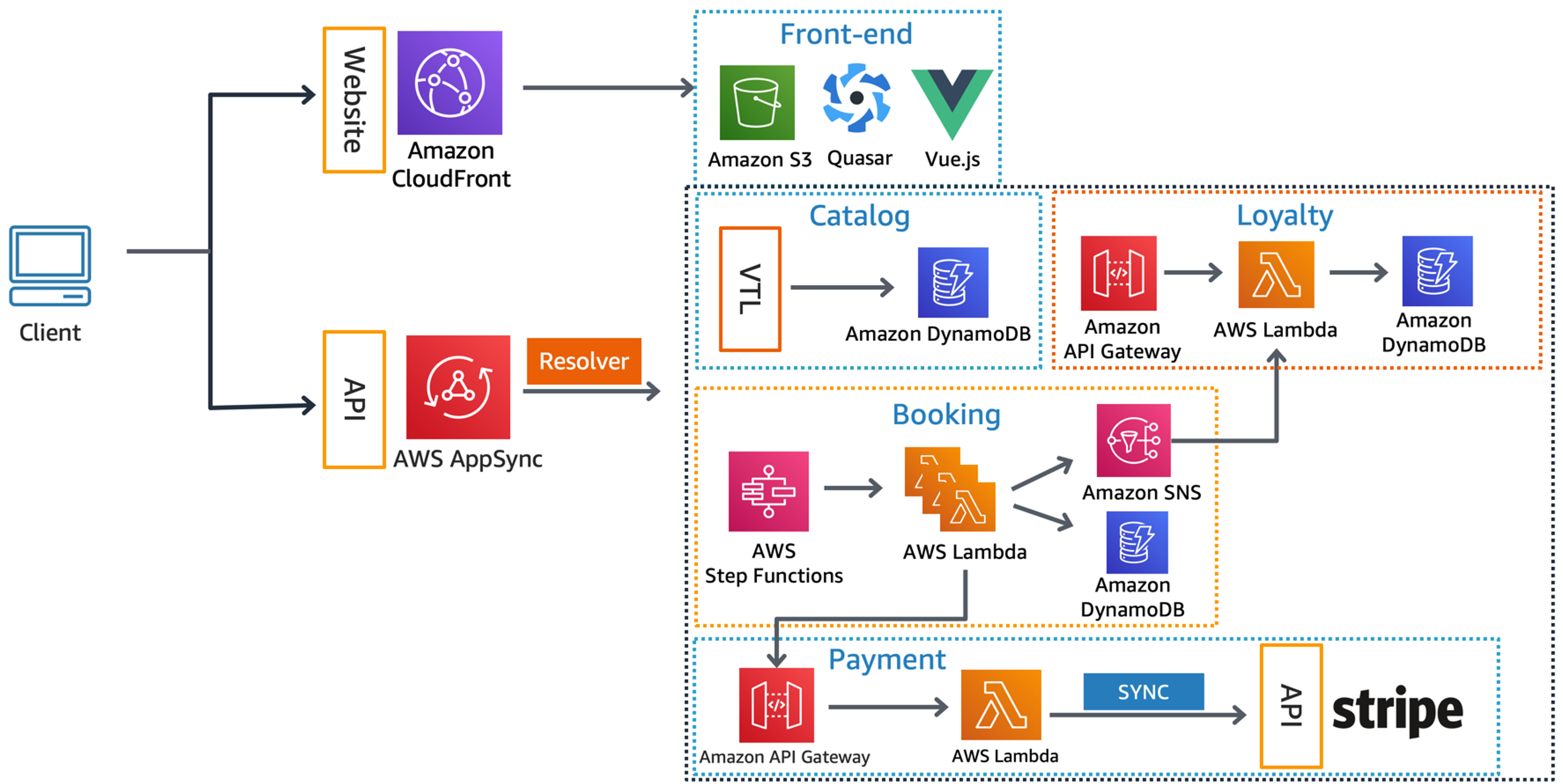}
    \caption{\app architecture.}
    \label{fig:airline_a}
\end{figure}

The \app application consists of several AWS serverless services: four DynamoDB tables, four S3 buckets, the SQS (queue) and SNS (notification) services, the AppSync API service, and 12 Lambda functions.
We utilized the AWS CloudTrail service to record the API activity of the application's storage and functions.

There are two main categories 
of business processes in the \app application's architecture, as shown in Figure~\ref{fig:airline_a}. 
The first category contains the \textbf{Catalog} and \textbf{Loyalty} processes: short-lived serverless functions that retrieve data, e.g., flight catalogs and loyalty points, from a DynamoDB table.
The second category includes the \textbf{Booking} process.
This process is initiated by an API request to book a flight and triggers a series of data access operations and Lambda function invocations in order to complete a flight booking transaction.
The booking process is data-driven and is dictated by the user's previous action, like choosing available seats on a selected flight, viewing their loyalty point balance, changing an existing booking, etc.

\textbf{Testbed and Simulation - }
To assess our \name stack, we built a cloud-based testbed.
Our testbed enables us to run a serverless application (e.g., \app) in a controlled environment and monitor the application's execution by collecting the activity logs.
In order to emulate benign and anomalous activity in the testbed, we ran a simulation script with multiple application use scenarios.
We configured AWS CloudTrail logging and activated event data logging on all of the cloud resources.
The Lambda function's invocations and execution logging logs were collected and merged using event-unique identifiers (requestID,EventID), and all data API access logs in the application were also recorded (see \ref{background_logging}).\label{subsec:simulation}
Simulations and recordings of the serverless application were performed by creating Python scripts that activate functions logically, just as a typical application user (client or customer) would.
In addition, errors that might occur if the user was not well-versed in the application were also simulated.
Then, we implemented several attacks by modifying the logic of some Lambda functions.
The simulation starts when the script (simulating a user) opens the browser with the address of the application and randomly performs one of the following actions: booking a flight, checking the loyalty point balance, or checking existing flight bookings.
Each action triggers a different business process and can also result in different action sequences (e.g., a booking could be one-way or round-trip).

\textbf{Attack Scenarios} - 
In our simulations, the application primarily operated in the benign mode, with the anomaly mode (simulating attacks) activated in about 2\% of runs. We simulated DoW and data leakage attacks, given their prominence as the top risks in serverless environments\cite{marin2022serverless, OWASP,csa2022toptthreat9}.
\begin{enumerate}[noitemsep,nolistsep]
    \item \textbf{DoW:} For the DoW attack simulation (see \ref{background_dow}), we altered the reserve booking Lambda function to execute numerous dummy API requests to a DynamoDB table, aiming to deplete the allocated budget.
    \item \textbf{Data Leakage:} For the data leakage simulation (see \ref{background_dl}), we modified the confirm booking Lambda function to direct data to a public bucket instead of private storage.
\end{enumerate}

\section{\label{sec:framework}\name Stack}
\begin{figure}[h]
  \centering
  \includegraphics[width=0.5\linewidth]{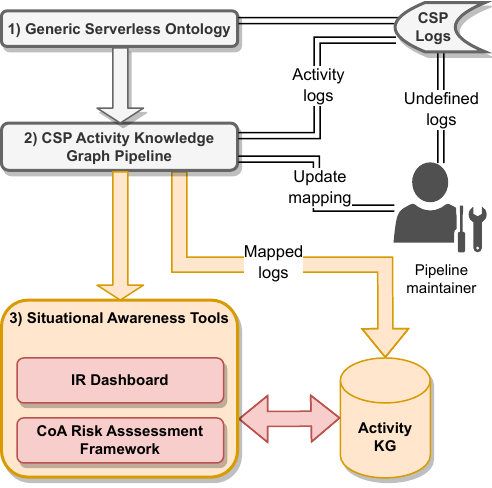}
  \caption{\name stack.}
  \label{fig:framework_stack}
\end{figure}

In this section, we present the \name stack, designed to enhance cybersecurity situational awareness in fully managed serverless applications.
Our approach begins by logging the activity of a specific CSP, such as AWS, via a native cloud audit service (e.g., CloudTrail in AWS), converting these logs into CSV format.
Our stack, as depicted in Figure~\ref{fig:framework_stack}, begins with the \emph{Serverless Ontology} layer (see Section~\ref{subsec:onto}), in which ontological entities and relationships within serverless applications are defined.
In the second layer of the stack, the \emph{CSP Activity Knowledge Graph} layer, the generic serverless ontology is extended for a specific CSP, such as AWS (see Section~\ref{subsec:activity_kg}), based on the activity logs collected.
In the \emph{AWS Application Activity Knowledge Graph} layer, the CSP-specific ontology is applied to the application's activity logs to generate the knowledge graph triplets using the \name pipeline.
These triplets, representing both implicit and explicit relationships, are stored in the Neo4j graph database\footnote{\url{https://www.neo4j.com/}}, facilitating efficient querying and management.
Once the application logs are processed and stored in the graph representation, the \emph{Situational Awareness} layer utilizes the generated knowledge graph for cybersecurity situational awareness (CSA) tasks.
In this paper, we demonstrate two CSA capabilities:
1. Incident Response (IR) dashboard (see Section~\ref{subsec:dashboard}), 
a custom-built dashboard, developed using Dash\footnote{\url{https://dash.plotly.com/}} and integrated with Neo4j, enables effective exploration and investigation of application activities.
2.  \emph{Criticality of Asset risk assessment framework} (see Section \ref{subsec:coa}), which assigns a criticality score to each resource, enhancing the graph representation with contextual information.

\subsection{\label{subsec:onto}Serverless Ontology}
The development of a serverless ontology involves the identification of activity (explicit) and contextual (implicit) entities and relationships within application activity logs in order to generate a structured, graph-based representation of application behavior.
To produce such a representation, the ontology map diverse application components, including compute, storage, and application resources, along with their roles and permissions.
This information is recorded in the activity logs, including details about resources interacting with one another and the permissions being used.

\begin{figure}[h]
    \centering
    \includegraphics[width=\columnwidth]{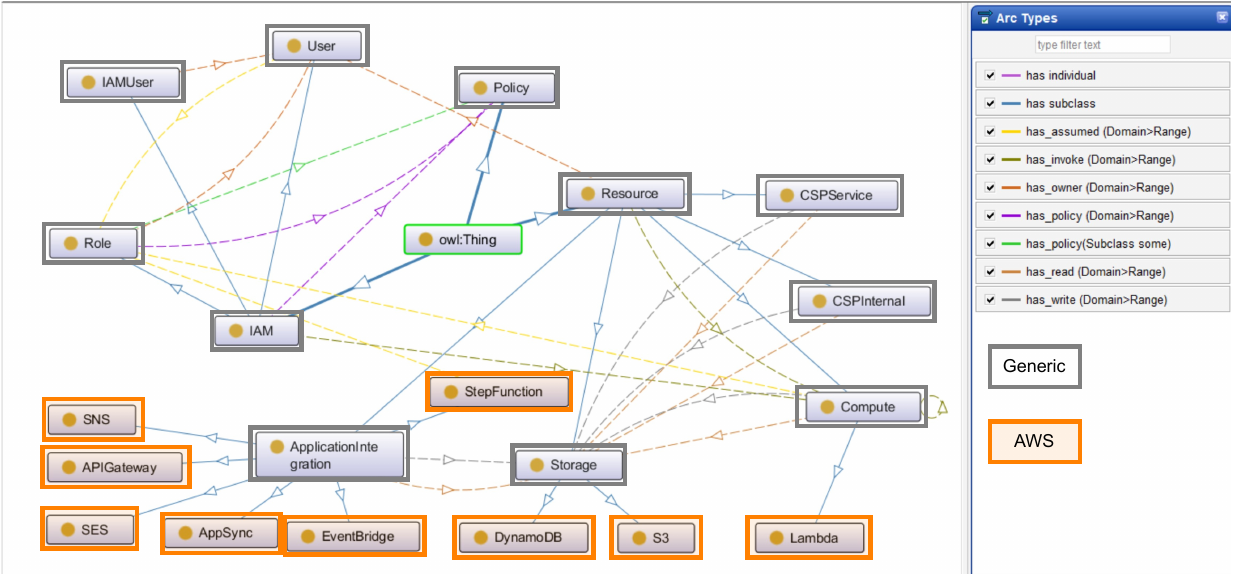}
    \caption{Visualization of the \name ontology  file.}
    \label{fig:ontology_static_rel}
\end{figure}

To define the ontology entities and their relationships, we employed the Owlready2 Python package~\cite{lamy2017owlready}, an implementation of the Ontology Web Language (OWL), and utilized Protege\footnote{\url{https://protege.stanford.edu/}} for visualization. The resulting \name ontology is presented in Figure~\ref{fig:ontology_static_rel}.
The serverless ontology includes the mapping of generic managed serverless application entities (in gray) to AWS serverless environment entities (in orange).
The solid blue lines represent inheritance relationships between the entities, denoted by the "has\_subclass" property.
In contrast, the dashed lines represent implicit relationships between the entities.
For example, the storage class represents the managed data sources and is related to the resource entities through the following implicit relationships: 1) it is a subclass of the resource entity, 2) it has an owner (inherited from the resource class), and (3) it implicitly indicates read/write access from other resources.

\textbf{Ontology Entities} - Application resources cover a variety of serverless services, while identity access management (IAM) identities denote the roles and users associated with a specific application.
As depicted in Figure~\ref{fig:ontology_static_rel}, the IAM identities are:
\begin{enumerate}[noitemsep,leftmargin=*]
    \item \textbf{User} - This entity signifies the root account name or ID, denoting the application's owner.
    \item \textbf{Role} - defines a set of permissions and access policies, which can be assumed by authorized users or resources to perform certain tasks within the cloud environment.
    \item \textbf{IAMUser} - a managed IAM identity created by a user capable of switching between roles (with different permissions) to perform various tasks.
    \item \textbf{Policy} - denotes policy permissions (e.g., PutObject used by a function to store an object in a storage service).
\end{enumerate}

Essentially, every application resource is an instance of a serverless service, and its owner is the user who created it.
When mapping each service, we created abstract categories to group services with a common activity log representation.
For instance, the ApplicationIntegration category represents multiple logic/rule-based serverless services, such as AWS Simple Notification Service (SES), EventBridge, and StepFunction (flow management). 
These services, which can access storage (inherited from the ApplicationIntegration entity) and invoke serverless compute services, are owned by the user (both are inherited from the resource entity).
Additionally, we created two more categories to handle other situations:
(1) CSPInternal, when the log entry does not contain enough information (e.g, internal logs in AWS); and (2) CSPService, when the log entry format is new or unmapped (does not exist in the ontology).
The serverless resource entity includes the following categories (see Figure~\ref{fig:ontology_static_rel}):
\begin{enumerate}[noitemsep,nolistsep]
    \item Compute - serverless functions.
    \item Storage - the application data and databases.
    \item ApplicationIntegration - the managed application building blocks.
    \item CSPService - represents new, unmapped service instances in the activity logs.
    \item CSPInternal - used as a placeholder for unmapped CSP environment service entities (e.g., internal trigger rule or scheduled event).
\end{enumerate}

\begin{figure*}[h]
    \centering
    \includegraphics[width=\textwidth]{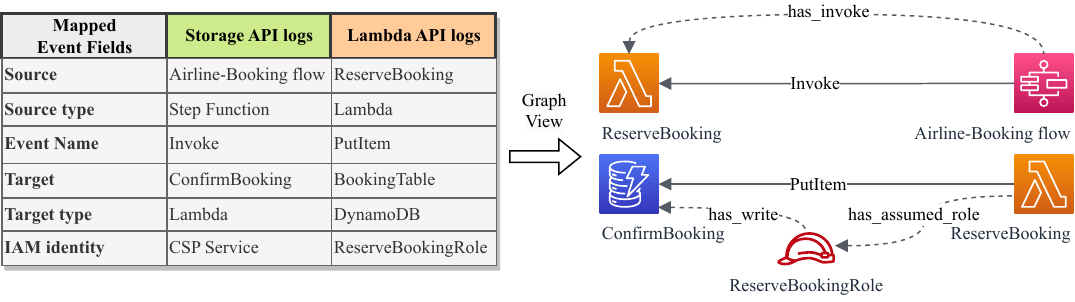}
    \caption{AWS log source types mapped and transformed using the \name ontology.}
    \label{fig:onto_demo}
\end{figure*}

\textbf{Explicit Relationships}  - Explicit relationships represent the interaction between cloud resources as an event relationship (e.g., A performed an API event on B).
Therefore, each API event log entry is translated into an explicit relationship where its attribute contains the event information (event name, source IP, region, etc.).
For example, as depicted in Figure~\ref{fig:onto_demo}, the ReserveBooking function performs a \textit{PutItem} event on the ConfirmBooking table; the function and DynamoDB table represent the explicit entities, and the \textit{PutItem} action is the explicit event relationship involving them.

\textbf{Implicit Relationships}  - Implicit relationships represent the additional information that can be retrieved from the logs (IAM identity and permission use).
These ontological relationships of each activity help create a context overlay on the application activity.
The implicit entity can be an IAM resource in use (e.g., a function assuming an IAM role to access another resource) or a synthetic entity (e.g., a step function invokes another function).
The implicit relationships are presented in the legend of Figure~\ref{fig:ontology_static_rel} and defined as follows:
\begin{enumerate}[noitemsep,leftmargin=*]
    \item has\_subclass - an ontological inheritance relationship.
    \item has\_assumed - a data property indicating the use of IAM roles by resources.
    \item has\_invoke - a data property denoting the invocation of an application function (e.g., invoking a Lambda function instance).
    \item has\_owner - a data property signifying the linkage between resources and their owners.
    \item has\_policy - a data property representing a synthetic relationship to the policies in use (a policy is a synthetic entity).
    \item has\_read/has\_write - a data property distinguishing read or write access to storage services.
\end{enumerate}

The nodes correspond to the application resource entities and IAM identities, and the edges denote the activity log (explicit) or contextual (implicit) relationships between the nodes.
Specifically, when generating the ontology, there is a need to map each log source format and define the corresponding application resource entities and IAM identity information in the ontology.
For example, certain actions, such as a function accessing a storage table, require resources to 'assume' an IAM role. Conversely, core serverless services, like flow management, queue, and notification services, inherently include policies upon application deployment.
To accurately reflect these dynamics, we model the IAM role as an implicit entity within our framework.
This distinction allows for a clear separation between the resources that assume IAM roles and the roles themselves, ensuring a more precise representation of IAM usage across different services.

In Figure~\ref{fig:onto_demo}, the table on the left contains two types of application activity log sources, storage API logs and Lambda API calls, respectively representing the function invocation and storage access activity.
The table contains all the necessary fields we mapped to create the activity graph representation on the right.
The activity log mappings are as follows:
\begin{enumerate}[noitemsep,nolistsep,leftmargin=*]
    \item \textbf{Invoke} event - This item represents the relationship where the serverless flow management service, \textit{Airline-Booking flow} (step function), invokes the serverless function, \textit{ConfirmBooking} (Lambda), as part of the application's operation.
    In the Lambda invocation event log, the Airline Booking flow uses a temporal role to access the function, but it is allowed to do so as part of the permissions of this service instance.
    \item \textbf{PutItem} - This event represents a data access performed by a function. 
    The ReserveBooking function must use the role assigned to it to perform actions.
    The function assumed by the role provided to it, ConfirmBookingRole, is to access the BookingTable DynamoDB instance.
    In the PutItem event, the Lambda function uses an assumed role and is connected to the implicit IAM Role entity with a \textit{has\textunderscore assumed} relationship.
    The role is connected to the target resource (the ConfirmBooking table), with the \textit{has\textunderscore policy} implicit relationship and representing the permission.
    Additionally, to distinguish write/read actions to each storage resource, an implicit relationship of \textit{has\textunderscore write}/\textit{has\textunderscore read} is added.
\end{enumerate}
The proposed ontology, based on generic managed serverless activity logs, enables us to create a parser to transform the semi-structured activity logs into an activity knowledge graph.

\subsection{\label{subsec:activity_kg}Activity Knowledge Graph Pipeline}
This subsection describes the \name pipeline used to create the application activity knowledge graph (KG).
The proposed pipeline is comprised of eight modules (A - H), which are presented in Figure~\ref{fig:Proposedpipeline} and used to create the activity KG.
We use the ontology (see Section~\ref{subsec:onto}) to perform two tasks:
(1) CSP-specific ontology construction and maintenance, and (2) KG construction.

\textbf{CSP-Specific Ontology Construction} - Since the \name ontology serves as an abstraction of serverless services, to apply its representation to new CSP activity logs, there is a need to (\textbf{A}) examine each new log source event format and update the ontology with new or revised service entities and relationships.
After the ontology has been updated, the new or modified parsing process must be incorporated into the mapping module (\textbf{D}) and the update procedure module (\textbf{E}).

\textbf{Knowledge Graph Construction} - Initially, the data from each log source is imported (\textbf{A}) and mapped in the graph-based representation (\textbf{D}).
The mapping process parses the logs into graph representation triplets (e.g., my-function, GetObject, my-bucket) which are inserted (\textbf{E}) in the activity KG database (\textbf{G}).
For each log event relationship, the activity KG stores the explicit relationship, referred to as an event relationship, and stores relevant attributes (e.g., date, user agent, error message, region).

To illustrate our methodology in the AWS environment, we extended the generic serverless ontology (from the first layer of the \name stack) relying on the \app logs, and implemented the \name pipeline (Figure~\ref{fig:Proposedpipeline}).
A step-by-step description of the \textit{PutItem} event depicted in  the mapping process in Figure~\ref{fig:onto_demo} is provided below:
\begin{enumerate}[noitemsep,nolistsep,leftmargin=*]
\item Extract the event name from the \textit{eventName} field.
\item Identify the source entity for the event using the \textit{userAgent} field.
This field provides information on the function environment, which enables the identification of the serverless function.
\item Identify the target resource using the \textit{eventSource} field, which indicates the event creator service type.
\item Derive the IAM entity and role name (if used) from the  \textit{userIdentity\_type} and  \textit{userIdentity\_principalId} fields, respectively.
\item Retrieve the target source storage name (DynamoDB table or S3 bucket) from the \textit{requestParameters\_tableName} or \textit{requestParameters\_bucketName} field, respectively.
\end{enumerate}

\subsection{\label{subsec:dashboard}CSA Tools: IR Dashboard}
\begin{figure*}[h!]
    \centering
    \includegraphics[trim={0 9.1cm 0 0},clip,width=0.99\textwidth]{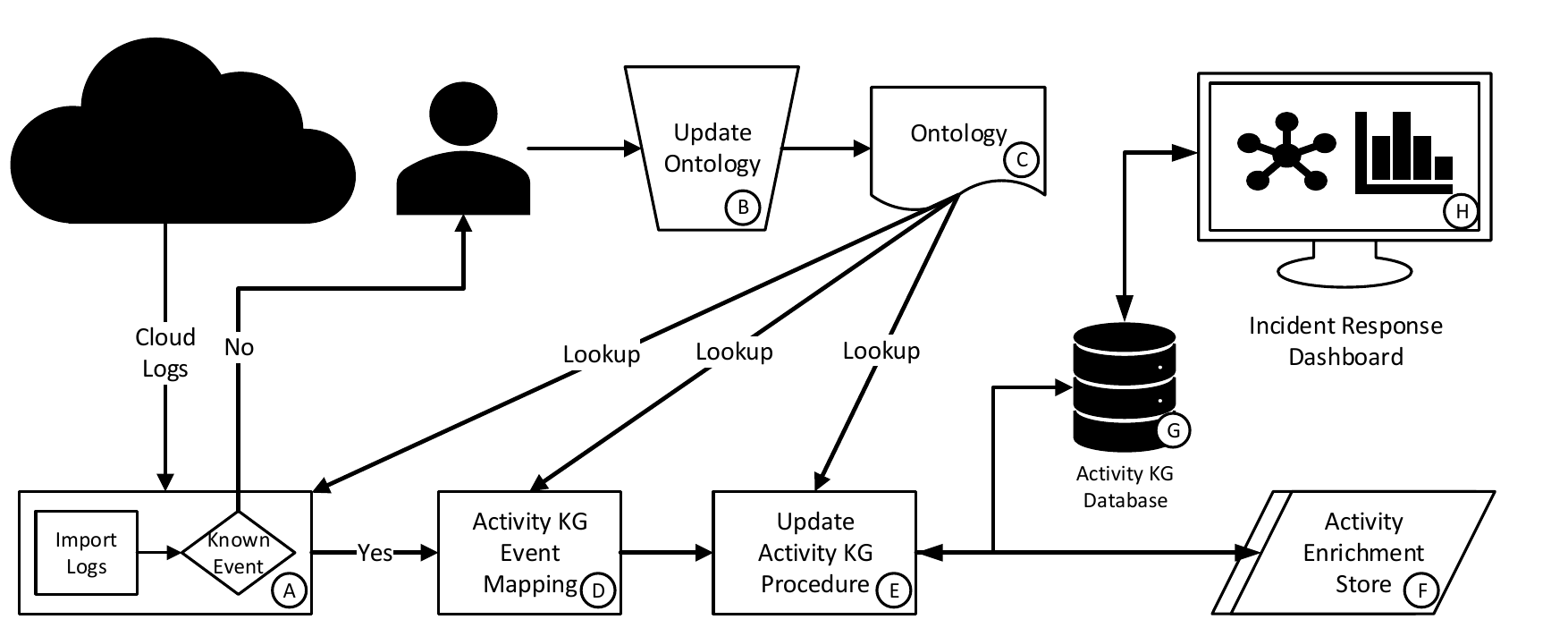}
    \caption{\name pipeline.}
    \label{fig:Proposedpipeline}
\end{figure*}

The dashboard (\textbf{H} in Figure \ref{fig:Proposedpipeline}) has a range of functionalities that enable security teams and developers to interact with the application activity graph representation.
Figure \ref{fig:dashboard_image} presents the \name dashboard's main screen.
As can be seen in the figure in (\textbf{A}), the application logs being examined define a time window (temporal) and can be adjusted by the user.
The type of node (resource) is indicated by icons, as shown in Figure \ref{fig:dashboard_image}, and the activity enrichment (confidentiality criticality of the asset ranking, described in Section \ref{subsec:coa}) is indicated by the background color of the node.
For example, C.1 is an AWS Lambda function, and its color indicates that its criticality rank is green, i.e., low confidentiality criticality.
A more specific search can also be performed via a unique API requestID search, as shown using the B filter option and marked in red (C.2).
The resource information (e.g., name, type, criticality) is displayed when selecting a single node.
Events are represented by the edges, and the edge thickness indicates the number of events logged within the selected time window. 
The raw event logs are displayed on the right side of the dashboard (frame D), and the distribution of the event type is presented below the graph activity (frame C.3).
Both of the events in frame D and the event distribution plot in C.3 are comprised of the logs of the selected nodes in frame C in the selected time window A.
By using the dashboard and leveraging its filtering, search, and query capabilities, users can quickly identify patterns, dependencies, and anomalies in application activity, thereby improving alert investigation task performance.
These dashboard features include:
\begin{itemize}[noitemsep,nolistsep,leftmargin=*]
\item A graphical representation of application activity, promoting a comprehensive understanding of the system.
\item Time window filtering capabilities, enabling the examination of data within specific time windows.
\item Resource filtering options, allowing users to focus on relevant components within the system.
\item Resource search functionality, facilitating rapid identification of specific resources.
\item Specific relationship search features, enabling analysis of connections between resources.
\item Presentation of raw log data, providing users with the most detailed information available.
\item Query execution using the activity KG, supporting advanced data retrieval and analysis.
\item Display of the event distribution based on selected time windows and resources, providing a broader perspective on system events.
\item Generation of reports, including the current graph view configuration and logs, enabling findings to be documented and communicated.
\item Visualization-based enrichment of the graph view via enrichment modules (e.g., asset criticality ranking), enhancing the context of the displayed data.
\end{itemize}

\subsection{\label{subsec:coa}CSA Tools: Criticality of Asset (CoA) Risk Assessment}
In a serverless application, various resources like Lambda functions and storage services work together to achieve specific goals (e.g., booking a flight).
The sequence in which these resources are executed is known as an application flow.
Asset prioritization, based on the risk each resource poses to the application flow during a cyberattack, enhances cloud environment observability for cybersecurity teams.
The CoA ranking can further guide incident evaluations and targeted defense strategies, thereby improving CSA critical decision-making.

CoA ranking is a complex task influenced by multiple factors, including:
1) The role within the application flow: For example, a Lambda function called L1 inserts and extracts data using an S3 bucket called B1 (two assets).
If B1 is attacked, only the L1 Lambda function would be affected, since L1 interacts solely with B1.
However, attacking an S3 bucket accessed by multiple Lambda functions could impact those functions and consequently impact other application flows.
2) The sensitivity of the data: Sensitive data like credit card information naturally raises an asset's importance.
3) Dynamic activities: For instance, infrequent yet critical processes like bi-daily updates may hold greater significance than more frequent but less crucial activities like weather reporting.

In 2008, Saaty~\cite{saaty2008decision} proposed the analytic hierarchy process (AHP), a theory of measurement through pairwise comparisons used to derive relative priorities in decision-making.
In these comparisons, experts use a scale of absolute judgments to determine how much one element dominates another based on a particular attribute or context.
In recent years, AHP has been applied in various domains, including evaluating the factors affecting blockchain~\cite{orji2020evaluating} and evaluating cyber attacks on machine learning production systems~\cite{bitton2023evaluating}.
In this study, we introduce an expert-based CoA ranking framework that employs both the AHP methodology and the proposed \name ontology to rank assets (resources) in the context of serverless application cybersecurity.

Our expert-based tool decomposes the challenging problem of asset prioritization into smaller sub-problems by generating a two-level hierarchy that includes the business process and asset type levels.
At the higher level, the business process level, the resources of a given application are assigned to their business processes.
In the lower level, the asset type level, each resource is categorized by its type (e.g., DynamoDB table) using the \name ontology. 
Then, AHP is used to perform pairwise comparisons between the resources at each level.
The use of the hierarchical structure in AHP reduces the number of pairwise comparisons required. 
In a flat structure, \(C_{\text{Naive}} = \frac{n(n - 1)}{2}\) comparisons are required for \(n\) resources.
With hierarchy, resources are divided into \(k\) subgroups.
Each subgroup \(i\) with \(m_i\) entities requires \(C_{\text{group}_i} = \frac{m_i(m_i - 1)}{2}\) comparisons.
The total becomes \(C_{\text{AHP}} = \sum_{i=1}^{k} C_{\text{group}_i}\).
Since \(m_i < n\), and the sum of all \(m_i\) is \(n\), \(C_{\text{AHP}}\) is less than \(C_{\text{Naive}}\) when \(k > 1\), the process becomes more efficient.
Finally, to ensure the validity of the prioritization process, the internal consistency of each expert ranking is examined.
To demonstrate the proposed method, we use the following example: Assume an application A, which consists of eight Lambda functions (L1-L8) and seven S3 buckets (S1-S7). 

The CoA ranking framework utilizes the \name ontology in order to solve the following challenge:
When comparing two different serverless resources, it is difficult to decisively evaluate which of the resources should be ranked as more critical.
For example, exploiting a Lambda function could compromise availability by disabling the app, whereas misusing an S3 bucket could lead to data leakage, affecting confidentiality. 
Comparing these two security scenarios on a large scale presents challenges, as each type of resource is associated with its unique set of threats and contextual factors.
To address the  complexities in this example, we employed our \name ontology to categorize the assets within each business process into disjoint groups based on type, such as Lambda functions, API gateways, or storage.

After grouping the resources in each business process by type, cybersecurity experts perform a pairwise comparison in the context of one component of the CIA triad (confidentiality, integrity, availability) as follows: 
First, a pairwise comparison is performed between the identified business processes using a scale of one to nine
where a score of one indicates that the first entity is the most critically important, a score of nine indicates that the second entity is the most critically important, and a score of five indicates that the two entities have equal importance.
 
For example, Table \ref{tab:ahp_calc_example} presents three business processes: B1, B2, and B3. 
Each business process contains two types of serverless services, denoted as G1 to G6. 
Within each of these groups, resources of the same type are clustered together. 
For instance, G1 includes resources L1, L2, and L3, while G2 is comprised of S1 and S2. 
In this terminology, 'B' stands for Business Process, 'G' represents a resource group as categorized by the \name ontology, 'L' denotes Lambda function resources, and 'S' signifies S3 resources. 
The ranking process involves pairwise comparisons at each level of the hierarchy. 
To illustrate, let us consider the final ranking score for resource S1, which is part of storage type G2 in business process B1. 
Initially, the ranking score for the business process level is calculated, yielding a normalized score of 0.7 for B1. 
Then, a pairwise comparison between G1 and G2 is performed within the context of B1, resulting in a score of 0.4 for G2. 
Finally, at the resource level, a comparison between S1 and S2 is made, with S1 receiving a score of 0.8. 
The final ranking score for S1 is then computed by multiplying these hierarchical scores together: \(0.7 \times 0.4 \times 0.8 = 0.224\).

After collecting all of the rankings from the annotators, we validate the internal consistency of each annotator ranking using the consistency ratio (CR).
According to the AHP, CR values that are lower than 0.1 are considered consistent.
Therefore, if the CR score value exceeds 0.1, the annotators are requested to update their ranking to maintain consistency.
By the end of this step, each annotator should have a CR value that is less than 0.1 for all comparison levels.
This validation is performed dynamically during ranking, using the CR measure.

Once the rankings of all of the experts have a valid CR, we examine the external agreement among different annotators.
This evaluation is performed using Kendall’s W test (i.e., Kendall’s coefficient of concordance), which is a nonparametric statistical test commonly used for assessing agreement among raters.
Kendall’s W ranges from zero (no agreement) to one (complete agreement), where a value over 0.6 is considered strong agreement~\cite{genest2011estimators,kendall1940method}.
Finally, we calculate the average score of each element at its respective level.
This hierarchy ranking is performed separately for each CIA context.

 \begin{figure}[h]
    \centering
    \includegraphics[trim=0cm 0cm 0cm 0cm, clip,width=\linewidth]{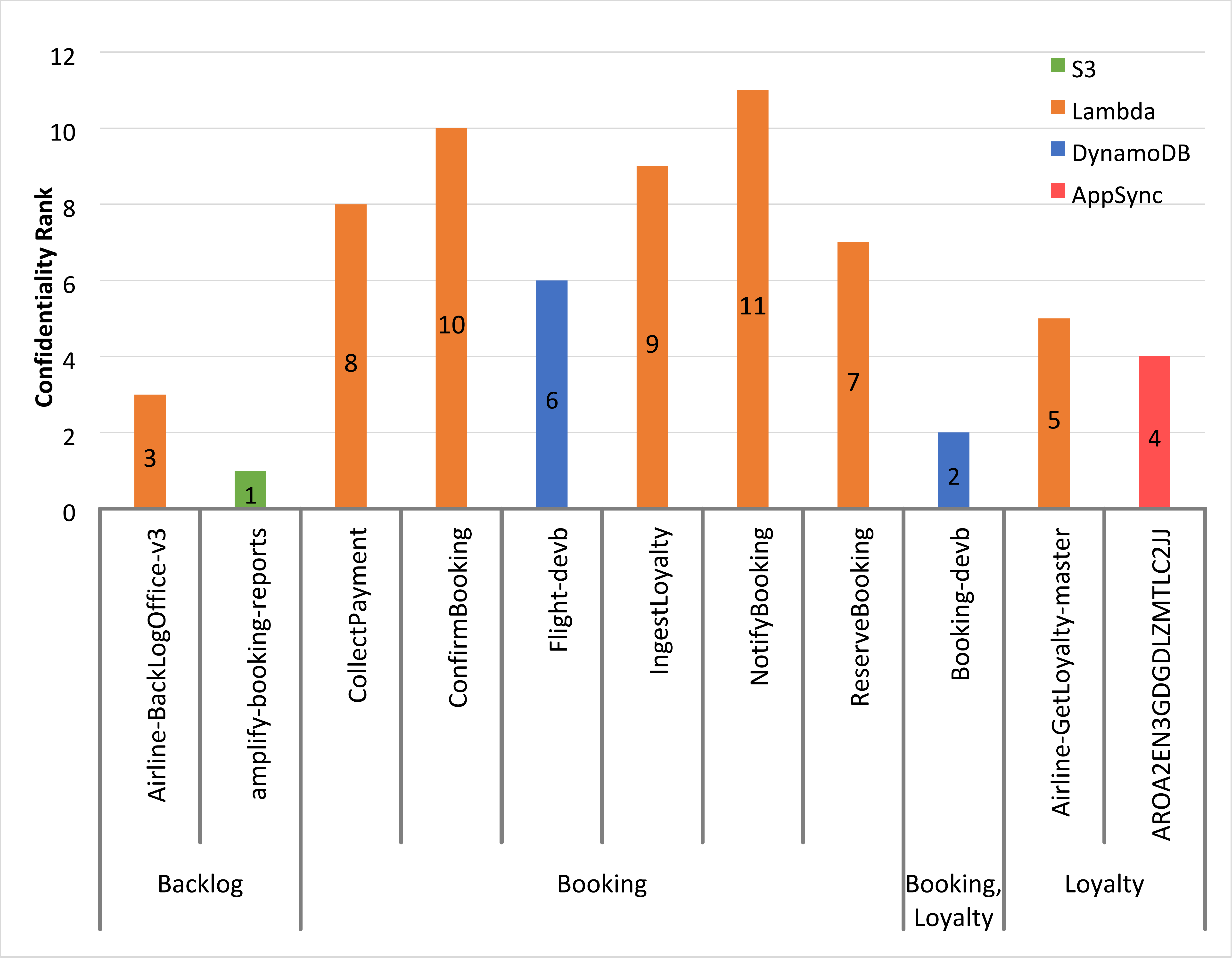}
    \caption{Criticality of assets: \app confidentiality ranking. On the x-axis, labels for processes are horizontal, and labels for resource names are vertical.}
    \label{fig:ahp_res}
\end{figure}

\begin{table*}[h]
\resizebox{\textwidth}{!}{%
\begin{tabular}{|
>{\columncolor[HTML]{C0C0C0}}c |ccccc|ccccc|ccccc|}
\hline
Bussiness level &
  \multicolumn{5}{c|}{B1(0.7)} &
  \multicolumn{5}{c|}{B2(0.15)} &
  \multicolumn{5}{c|}{B3(0.15)} \\ \hline
Resource type &
  \multicolumn{3}{c|}{G1(0.6)} &
  \multicolumn{2}{c|}{G2(0.4)} &
  \multicolumn{3}{c|}{G3(0.1)} &
  \multicolumn{2}{c|}{G4(0.9)} &
  \multicolumn{2}{c|}{G5(0.5)} &
  \multicolumn{3}{c|}{G6(0.5)} \\ \hline
Resource &
  \multicolumn{1}{c|}{L1(0.6)} &
  \multicolumn{1}{c|}{L2(0.2)} &
  \multicolumn{1}{c|}{L3(0.2)} &
  \multicolumn{1}{c|}{S1(0.8)} &
  S2(0.2) &
  \multicolumn{1}{c|}{L4(0.4)} &
  \multicolumn{1}{c|}{L5(0.4)} &
  \multicolumn{1}{c|}{L6(0.2)} &
  \multicolumn{1}{c|}{S3(0.7)} &
  S4(0.3) &
  \multicolumn{1}{c|}{L7(0.8)} &
  \multicolumn{1}{c|}{L8(0.2)} &
  \multicolumn{1}{c|}{S5(0.4)} &
  \multicolumn{1}{c|}{S6(0.3)} &
  S7(0.3) \\ \hline
Rank &
  \multicolumn{1}{c|}{0.252} &
  \multicolumn{1}{c|}{0.084} &
  \multicolumn{1}{c|}{0.084} &
  \multicolumn{1}{c|}{0.224} &
  0.056 &
  \multicolumn{1}{c|}{0.006} &
  \multicolumn{1}{c|}{0.006} &
  \multicolumn{1}{c|}{0.003} &
  \multicolumn{1}{c|}{0.0945} &
  0.0405 &
  \multicolumn{1}{c|}{0.06} &
  \multicolumn{1}{c|}{0.015} &
  \multicolumn{1}{c|}{0.03} &
  \multicolumn{1}{c|}{0.0225} &
  0.0225 \\ \hline
\end{tabular}
}
\caption{AHP rank calculation example.}
\label{tab:ahp_calc_example}
\end{table*}
\section{\label{sec:user_study}
Incident Response Dashboard User Study}
We performed a user study involving manual log examination and analysis.
Thirty-nine participants took part in the user study: 28 male and 11 female computer engineering students (25 B.Sc. students and 14 M.Sc. students).
The experiment was approved by the Institutional Review Board (IRB).
First, a questionnaire was used to assess the participants' prior knowledge and expertise in serverless computing, cloud computing, and cybersecurity log investigation. 
The participants rated their expertise on a scale of one to five; the average score was below 2.6, with a variance of less than one for all questions. 
This indicates limited experience in cloud environments, which allows us to demonstrate the potential of using the proposed dashboard for the common user.
To ensure that participants were adequately prepared, training sessions were held to familiarize participants with the experimental protocol, \name dashboard, and the baseline tools.
In the user study, we measured the time it took for the participants to perform IR (log filtration, event context detection, and activity aggregation) and their accuracy in performing these tasks during the experiment.
\subsection{Experimental Protocol}
Before the experiment, a 25-minute presentation was delivered to familiarize the participants with the serverless environment, its services, and API audit logs, and demonstrate the functionality of a serverless application (\app) and the cybersecurity incidents they will encounter in the experiment.

Following the presentation, each participant was provided with 15 minutes of hands-on training on one of the tools utilized in the study (either the baseline tool or the \name dashboard). 
This was done to ensure that the participants were proficient in performing the IR tasks.
The training was designed in the form of do-it-yourself (DIY) missions that reflect the IR tasks examined in the study.
After the hands-on training, participants were asked to complete an online IR questionnaire as they performed five IR tasks in two simulated cybersecurity scenarios: data leakage and DoW. 
The duration of the experiment was 80 minutes.

During the experiment, in order to measure the time it took the participants to perform the investigation task, we created a briefing screen.
On the briefing screen, the participants were not permitted to use the tool but were allowed to read about the investigation task.
This screen appeared before each task and presented information about the task but did not include fields for answers.
The information presented included the goal of the task, instructions on how to approach it, and other relevant information.
When the participant navigated to the next page, the timer was initialized, and the participant was allowed to access the tool and complete the questionnaire.
Since each investigation task was divided into subtasks, knowledge of the correct response for each subtask was crucial to moving forward and completing the task.
Therefore, for each task, the correct response for the previous tasks was provided to minimize trailing errors.

To evaluate the IR dashboard's effectiveness in log analysis, two scenarios were chosen for their ability to represent distinct techniques. 
The first scenario (data leakage) requires isolating data access of a specific function and offers a unique investigative angle, requiring a context-aware log analysis strategy for problem resolution in order to resolve.
Conversely, the second scenario (denial-of-wallet) is tailored to reflect typical risk situations in serverless systems, focusing on identifying targeted attacks against a resource using a burst of traffic (reflected in a burst of API activity).

\textbf{Data Leakage Scenario} - In this scenario, participants investigated an alerted serverless function that performed unwanted access to private storage.
The compromised function, designed to work with public storage, was manipulated to gain access to sensitive data, resulting in a potential data leak.
The investigation process for the data leakage scenario involved the following steps that were explained to the participant:
\begin{enumerate*}
    \item Locating the alert event using the unique event identifier and extracting the resources and event name.
    \item Counting the number of resources that interacted with the extracted resources in two different time windows.
    \item Determining the storage (name and type) in which the alerted function was accessed during the two time windows.
\end{enumerate*}

\textbf{DoW Attack Scenario} - In this scenario, participants were tasked with investigating a given function in the \app application.
The targeted function was observed to repeatedly make API calls to access a data source, resulting in an increase in both data transmission and execution time.
The investigation process for the DoW attack scenario consisted of the following steps that were explained to the participant:
\begin{enumerate*}
    \item Identifying the suspected function using the unique event identifier provided in the alert and retrieving all relevant function interactions.
    \item Evaluating the interaction volume across three time windows to determine if the increased volume of function invocations indicates a DoW attack. 
\end{enumerate*}

\textbf{Comparison Baseline Tool:}
While Amazon CloudWatch\footnote{\url{https://aws.amazon.com/cloudwatch/}} is an integral tool for monitoring application operational health and performance, it primarily focuses on streamlining application activity, which is distinct from the granular log analysis often required for incident response. 
In contrast, the use of Microsoft Excel for log analysis is a common practice, and its use for this purpose is acknowledged by the security community, with endorsements by the SANS Institute for its adequacy in tasks such as data manipulation, filtration, aggregation, and visualization\cite{SANSWhitepaper}. 
Furthermore, several academic papers and industry recommendations substantiate the decision to utilize Excel as the baseline tool for analyzing and visualizing API event data. 
Rohde et al.\cite{rohde2023using} explored the use of Excel in logging and evaluating public displays, demonstrating its versatility across different application domains, while Awaysheh et al.\cite{awaysheh2020next} highlighted Excel's capabilities in federating big data access control, emphasizing its robustness in handling complex data structures. 
Excel's efficacy in assessing security risks in the Internet of things was discussed by Abbass et al.\cite{abbass2019assessing}, further validating its utility in analyzing API event data. 
Excel's intuitive interface and robust features, including filtration, aggregation, and visualization capabilities comparable to other log analysis tools, make it commonly used in both academic and practical contexts, influencing its selection for our user study.

\subsection{Results}
In the experiment, participants utilizing the \name dashboard demonstrated superior performance compared to those using the baseline in terms of the two examined KPIs: MTTD and accuracy of task results.

\begin{table}[h]
    \centering
    \begin{tabularx}{\columnwidth}{l|XX|XX}
        \hline
        \multirow{2}{*}{Task} & \multicolumn{2}{c|}{Baseline} & \multicolumn{2}{c}{\name Dashboard} \\
        & Accuracy & Time & Accuracy & Time \\
        \hline
        T1 & 0.843 & 7.513 & \textbf{0.933} & \textbf{4.063} \\
        T2 & \textbf{0.661} & 24.841 & 0.653 & \textbf{4.382} \\
        T3 & 0.771 & 6.529 & \textbf{0.956} & \textbf{3.663} \\
        T4 & 0.764 & 12.990 & \textbf{0.867} & \textbf{5.387} \\
        T5 & 0.821 & 7.442 & \textbf{0.954} & \textbf{5.712} \\
        \hline
    \end{tabularx}
    \caption{Mean accuracy score and mean time to complete each task (in minutes).
    For each task, bold values indicate better performance.}
    \label{tab:mean_score}
\end{table}

\begin{table}[h]
    \centering
    \begin{tabularx}{0.44\textwidth}{@{}l|X|X@{}} 
    \hline
     Task &  p-value &  t-statistic \\ 
    \hline
    Task 1 & -5.268 &  1.092e-05 \\
     Task 2 &  -6.950 &  1.016e-07 \\
     Task 3 &  -2.795  &  0.009  \\
     Task 4 &  -5.343 &  8.839e-06 \\
     Task 5 &  -1.645 &  0.110    \\
    \hline
    \end{tabularx}
    \caption{T-test results.}
    \label{tab:t_score}
\end{table}
During the experiment, we observed that training participants in the use of the baseline tool took more time than that of the dashboard, specifically in the use of constructing Excel pivot tables to aggregate the data; it took participants using the \name dashboard an average of 44.3 minutes to perform the tasks and complete the questionnaire compared to the 81.5 minutes required by those using the baseline.
We encouraged participants with little or no Excel Workbook (the baseline tool) experience to use pivot tables, as they only need to be set up once per log source and enable faster filtration and aggregation procedures.
A step-by-step tutorial was available during the experiment.
Following the experiment, the participants who used Excel reported that using the pivot table was not intuitive and took more time than they expected.
While tasks two and four, which involved creating pivot tables, took over twice as long with Excel compared to \name dashboard, tasks one, three, and five were quicker to complete using Excel, since participants could utilize the previously created pivot tables.

The accuracy and MTTD for each task are presented in Table~\ref{tab:mean_score}.
We note that the baseline tool outperformed the dashboard on just one task (task 2); however, the difference in performance is slight (0.008\%).
The dashboard outperformed the baseline tool on all the rest of the examined tasks with a maximum improvement in accuracy of 18.5\%.

When examining the MTTD, 
we note that the use of the dashboard reduced the MTTD on all tasks, with a reduction of 20 minutes seen for task 2, the only task that the baseline outperformed the proposed tool on. 
For all the examined tasks, the MTTD was significantly shorter when using the \name dashboard, proven statistically using a t-test~\ref{tab:t_score}.
Additionally, the average accuracy across tasks was higher for the proposed dashboard (0.873) compared to the baseline tool (0.772).
\section{Demonstration of the Criticality Assessment Ranking Framework}\label{subsec:ac_airline}
In this section, we demonstrate how the criticality of asset framework can be applied to an application to create a confidentiality risk score for each resource.

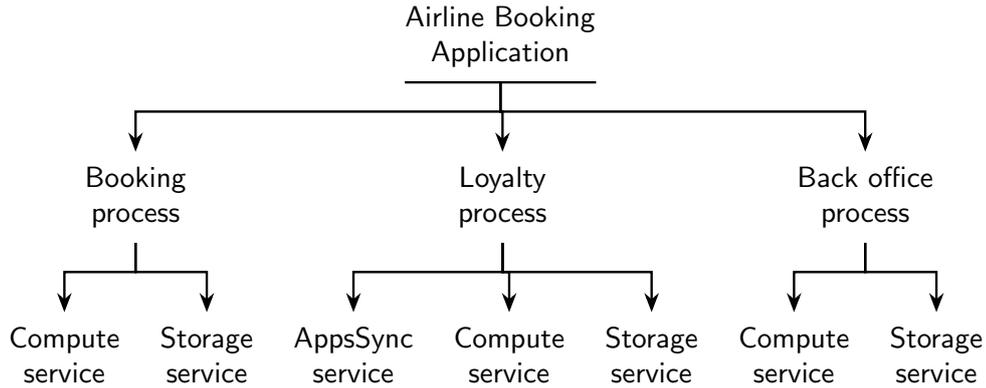
\begin{figure}[h]
\centering
\resizebox{\columnwidth}{!}{%
\begin{forest}
      for tree={
        align=center,
        parent anchor=south,
        child anchor=north,
        font=\sffamily,
        edge={thick, -{Stealth[]}},
        l sep+=12pt,
        edge path={
          \noexpand\path [draw, \forestoption{edge}] (!u.parent anchor) -- +(0,-10pt) -| (.child anchor)\forestoption{edge label};
        },
        if level=0{
          inner xsep=0pt,
          tikz={\draw [thick] (.south east) -- (.south west);}
        }{}
      }
      [Airline Booking\\Application
        [Booking\\process
            [Compute\\service]
            [Storage\\service]
            ]
        [Loyalty\\process
            [AppsSync\\service]
            [Compute\\service]
            [Storage\\service]
            ]
        [Back office\\process
            [Compute\\service]
            [Storage\\service]
            ]
      ]
    \end{forest}}
\caption{CoA framework -  \app hierarchy.}
\label{fig:AHPHi}
\end{figure}

\subsection{Analytical Hierarchy Process Construction}
We constructed the AHP levels (as depicted in Figure \ref{fig:AHPHi}) as follows:
(1) For the first level of the hierarchy, the business process level, we decomposed the Airline Booking application into its main business processes: booking flights, handling loyalty credits, and managing the back office.
The business processes were identified in the documentation provided by the application (see Section~\ref{sec:demoapp}).
(2) For the second level of the hierarchy, the resource type level, we used the \name ontology to determine how to group the \app resources into inner groups for each business process.

\subsection{Expert-Based Questionnaire}
Using the AHP hierarchy, we created a pairwise-based questionnaire.
Later, six annotators who implemented, simulated, and studied the application completed the questionnaire.
The context the experts used to weigh each comparison is the confidentiality of the application information.
The questionnaire's hierarchy was created, as described in Section~\ref{subsec:coa}, resulting.
In addition, among the experts, we obtained a Kendall-W score of 0.7179, which is considered strong agreement.
The ranking starts at one, a value which is assigned to the most critical resource in the context of the questionnaire.
In our evaluation, the \textit{back office} process was considered the most critical by all of the experts.
The second most critical process was the loyalty process, while the least critical process was the booking process which received the lowest rank. 
\subsection{Ranking Results}
The application resources ranking is presented in Figure~\ref{fig:ahp_res}, which shows that the most critical asset of the \app application is the \textit{amplify-booking-reports} bucket.
This result make sense, because that bucket is part of the \textit{Back office} business process, which handles the application data and backs up users' information.
The \textit{Loyalty} and \textit{Booking} processes were ranked evenly in terms of their weights, and their data criticality ranks were equal.
The process ranking impacts the final resource ranking, as the score is propagated by each level of the hierarchy.
This explains the relatively high rank of the \textit{Airline-BackLogOffice-v3} function, which was ranked as the most critical function and has access to the most critical storage instance, the \textit{amplify-booking-reports}.
\section{\label{sec:disscussion}Discussion}
Our proposed \name stack introduces a streamlined approach for cybersecurity in serverless architectures, leveraging native logging capabilities, such as AWS CloudTrail, for comprehensive monitoring with no additional overhead. 
\name stack's adaptability lies in its utilization of existing logs, ensuring seamless integration into existing workflows without impacting system performance.

However, a notable limitation is the requirement for ongoing maintenance to update the ontology and mapping processes in response to evolving log formats and new log data sources.
To overcome this, a maintenance strategy should be employed as part of \name stack's implementation.

Extending our method to accommodate other CSPs, such as Azure, requires some adjustments to the mapping process (see Section~\ref{subsec:activity_kg}). 
This adaptation, which is similar to maintaining mappings for CloudTrail, involves updating the \name pipeline (in \name stack this is performed by the pipeline maintainer, as can be seen in Figure 4), ensuring \name stack's applicability across various logging environments without imposing overhead on the application's operation.
Alternatively, instead of extending our proposed stack ontology and pipeline to other CSPs, one can utilize the proposed \name stack by converting each data source into the AWS log format (for example, convert Azure log analytics log format into CloudTrail).

\section{\label{sec:related}Related Work}

\begin{table*}[!h]
    \centering
    \label{tab:Comparison}
    \resizebox{\textwidth}{!}{%
    \begin{tabular}{|l|l|l|l|l|l|l|}
    \hline
    \multicolumn{1}{|c|}{\textbf{Method}} & \multicolumn{1}{c|}{\textbf{Year}} & \multicolumn{1}{c|}{\textbf{Purpose}} & \multicolumn{1}{c|}{\textbf{Environment}} & \multicolumn{1}{c|}{\textbf{\begin{tabular}[c]{@{}c@{}}Implementation\\ requirements\end{tabular}}} & \multicolumn{1}{c|}{\textbf{Overhead}} & \multicolumn{1}{c|}{\textbf{Dataset Provided }} \\
    \hline
    Trapeze~\cite{alpernas2018secure} & 2018 & Information flow control & AWS/OpenWhisk & Shim layer & Yes & No \\
    \hline
    SecLambda~\cite{jegan2020guarding} & 2020 & Information flow control & Kubernetes/AWS & Controller logic & Yes & No \\
    \hline
    Valve~\cite{datta2020valve} & 2020 & Information flow control & OpenFaaS & Controller logic & Yes & No \\
    \hline
    WILL.IAM~\cite{sankaran2020workflow} & 2020 & Information flow control & OpenFaaS & Modifying OpenFaaS & Yes & No \\
    \hline
    SCIFFS~\cite{polinsky2021sciffs} & 2021 & Information flow control & OpenFaaS & Modifying OpenFaaS & Yes & No \\
    \hline
    ALASTOR~\cite{datta2022alastor} & 2022 & Auditing framework & OpenFaaS & Modifying OpenFaaS & Yes & No \\
    \hline
    Kalium~\cite{jeganguarding} & 2023 & Control flow integrity & OpenFaaS & Modifying OpenFaaS & Yes & No \\
    \hline
    PrivFlow~\cite{satapathy2023disprotrack} & 2023 & Auditing framework & Docker & Shim layer & Yes & No \\
    \hline
    Our work & 2024 & Auditing framework & AWS & None (Plug \& play) & No & Yes \\
    \hline
    \end{tabular}%
    }
    \caption{Comparative analysis of serverless security frameworks.}
\end{table*}
\textbf{Cybersecurity Graph Representation}  - 
In recent years, graph-based models have gained traction in cybersecurity as a means of enhancing situational awareness. 
Esser \etal~\cite{esser2021multi} pioneered a multi-dimensional data model based on labeled property graphs using process logs (system logs).
Their model supports complex queries involving multiple entities.
Several other studies extended graph-based models for various cybersecurity applications. 
For example, Gove~\cite{gove2021automatic} and Böhm \etal~\cite{bohm2021visual} emphasized visualization, using network and system logs, respectively, while Brandón \etal~\cite{brandon2020graph} and Guo \etal~\cite{guo2020graph} applied graphs to managed microservices for trace and root cause analysis. 
Marco \etal~\cite{angelini2021bucephalus},
and Sanders \etal~\cite{sanders2019mining} explored the fields of visual analytics and access control through various graph representation techniques.
However, adapting these methods to serverless environments requires modifications to data models to accommodate new log formats and application workflows.
In addition, several studies leveraged KG representation \cite{mendsaikhan2020quantifying}, however they focused on different environment domains and applications (such as system-level loggings and CVE database management).

In contrast to the abovementioned studies, we focus on managed cloud-based environments, specifically serverless architectures, and consider the unique challenges they present for graph-based cybersecurity analytics.
Banse \etal~\cite{banse2021cloud} proposed the Cloud Property Graph, which is focused on a managed cloud-based environment and bridges static code analysis and runtime.
However, the authors relied on access to the application's cloud environment, whereas our proposed method relies solely on the application logs.

\textbf{Serverless Security Frameworks}  - 
Datta \etal~\cite{datta2020valve} introduced Valve, a framework that functions as a proxy for all requests and enables developers to exert complete fine-grained control of information flows in their applications.
In another study, Datta \etal~\cite{datta2022alastor} proposed ALASTOR, an audit framework that records serverless function activity in both the system and application layers (in unmanaged environments) to represent each function's behavior.
A study performed by \etal~\cite{alpernas2018secure} presented Trapeze, a shim layer for monitoring and tracking the global flow of information in the system by monitoring the inputs and outputs of all functions in the system. 
Sankaran \etal~\cite{sankaran2020workflow} presented WILL.IAM, a workflow-aware access control model and reference monitor that meets the functional requirements of the serverless computing paradigm; the authors compared WILL.IAM's performance to that of Valve and Trapeze and showed that their proposed method had greater computation overhead efficiency.
A security framework called SecLambda, which allows developers to run customized security tasks to protect their applications, supports flexible security policies and provides extended logging by using graphs to describe application flow, was suggested by Jegan \etal~\cite{jegan2020guarding}.
Polinsky \etal~\cite{polinsky2021sciffs} proposed decentralized information flow control (DIFC) and securing client information flows using function as a service (SCIFFS), demonstrating the use of queries to search for data exposure threats.
Obetz \etal~\cite{obetz2019static} proposed using code and configuration to construct a static call graph for AWS serverless applications to improve visibility.
Jegan \etal~\cite{jeganguarding} proposed a control flow integrity (CFI) method called Kalium, which has lower overhead and latency than Valve and Trapeze.
Farshchi \etal~\cite{farshchi2015anomaly} investigated anomaly detection in cloud application operations using log metric correlation, revealing challenges and limitations in system monitoring with logs.
Garg \etal~\cite{satapathy2023disprotrack} proposed PrivFlow, a workflow security framework applied on OpenFaas demo application; in their evaluation, the authors compared its performance to that of WILL.IAM~\cite{sankaran2020workflow} and found that PrivFlow reduced application overhead.

As seen in Table~\ref{tab:Comparison}, our proposed method distinguishes itself from existing serverless security frameworks in several ways.
In contrast to previous approaches that primarily function in unmanaged environments or require access to the application's environmental code, our method operates within managed cloud environments. 
It relies exclusively on the audit logs provided by managed environments (i.e., AWS CloudTrail).
Our approach eliminates operational overhead entirely, and it is application-agnostic.

\section{\label{sec:conc}Conclusions and Future Work}
This paper introduced the \name stack, a three-layer security scheme designed to address the unique cybersecurity challenges in fully managed serverless environments.
The scheme comprises a generic serverless ontology and a generic CSP activity knowledge graph.
In the third layer, we address the limitations of reduced observability and situational awareness capabilities by developing two situational awareness tools that leverage the graph-based representation: an incident response dashboard, and criticality of assets risk assessment framework.

Our comprehensive evaluation demonstrated the effectiveness of the \name stack.
Specifically, the IR dashboard led to a reduction in the MTTD by up to 18\%. 
Additionally, the average accuracy across tasks was higher for the proposed dashboard (0.873) compared to the baseline tool (0.772). 
The CoA risk assessment framework, operating within a confidentiality context, demonstrated a strong Kendall-W agreement score of 0.71 in resource prioritization.

The \name stack provides a scalable, adaptable foundation for developing other new cybersecurity tools.
It is particularly useful for security teams lacking application access (code, environment), enabling them to tailor the stack to meet specific needs and requirements.

Future research directions include the automation of log unification across multiple CSPs, extending situational awareness capabilities, and exploring graph-based representations for proactive threat hunting in managed serverless environments.

\bibliography{bibfile}
\end{document}